\documentclass[epj]{svjour}
\usepackage{graphics}

\usepackage{multicol}
\usepackage{graphicx}
\usepackage{booktabs}
\usepackage{amssymb, bm, mathrsfs, bbm, amscd}
\usepackage[tbtags]{amsmath}
\usepackage{lastpage}
\usepackage{changes}
\usepackage{dcolumn}
\usepackage{longtable}
\usepackage[numbers, sort&compress]{natbib}
\usepackage{graphicx}
\usepackage{dcolumn}
\usepackage{bm}
\usepackage{amsmath}
\usepackage{flushend}
\usepackage{changes}
\usepackage{multirow}
\usepackage{longtable}
\usepackage{threeparttable}
\usepackage{cancel}
\usepackage{float}

\begin{document}
\title{Systematic study of proton radioactivity of spherical proton emitters with Skyrme interactions}
\author{Jun-Hao Cheng\inst{1,2}\and Xiao Pan\inst{1}\and You-Tian Zou\inst{1}
\and Xiao-Hua Li\inst{1, 3, 4}\thanks{\emph{e-mail:} lixiaohuaphysics@126.com }
\and Zhen Zhang\inst{5, }\thanks{\emph{e-mail:} zhangzh275@mail.sysu.edu.cn }
\and Peng-Cheng Chu\inst{6, }\thanks{\emph{e-mail:} kyois@126.com }
}
%
%
\institute{School of Nuclear Science and Technology, University of South China, Hengyang 421001, China 
\and Nuclear Power Institute of China, Chengdu 610041, China
\and Cooperative Innovation Center for Nuclear Fuel Cycle Technology $\&$ Equipment, University of South China, Hengyang 421001, China 
\and Key Laboratory of Low Dimensional Quantum Structures and Quantum Control, Hunan Normal University, Changsha 410081, China 
\and Sino-French Institute of Nuclear Engineering and Technology, Sun Yat-sen University, Zhuhai 519082, China 
\and School of Science, Qingdao Technological University, Qingdao 266000, China}
\date{Received: date / Revised version: date}
%
\abstract{
Proton radioactivity is an important and common process of the natural radioactivity of proton-rich nuclei. In our previous work [J. H. Cheng et al., Nucl. Rhys. A \textbf{997}, 121717 (2020)], we systematically studied the proton radioactivity half-lives of $53\leq Z \leq83$ nuclei within the two-potential approach with Skyrme-Hartree-Fock. The calculations can well reproduce the experimental data of spherical nuclei. The present work is an extension of the previous work. We systematically study 32 proton radioactivity half-lives of spherical nuclei with the two-potential approach with Skyrme-Hartree-Fock (TPA-SHF) within 115 different versions of the Skyrme interactions. The calculated results indicate that the SLy7 Skyrme interaction gives the lowest r.m.s. deviation in the description of the experimental data of the spherical proton emitters among all the different versions of the Skyrme interactions. In addition, we use this model with the SLy7 Skyrme interaction to predict the proton radioactivity half-lives of 7 spherical nuclei in the same region, whose proton radioactivity is energetically allowed or has been observed but is not yet quantified. 
} 

\maketitle

\section{Introduction}
\label{section 1}

In 1970, proton radioactivity was firstly discovered by Jackson \emph{et al} \cite{JACKSON1970281} in the isomeric state of $^{53}$Co$^m$ which spontaneously decays to the ground state of $^{52}$Fe by emitting a proton. Since then, the proton radioactivity has been discovered in more and more nuclei in the protons number region $Z=51-83$ \cite{SONZOGNI20021}, and thus the study of proton radioactivity becomes a hot topic in nuclear physics \cite{PhysRevC.85.011303, Zdeb2016, Deng2019}. The systematic study of proton radioactivity is an important subject in relation to the extensive theoretical studies of proton dripline nuclei. As a typical decay mode of nuclei beyond the proton drip line, the proton radioactivity serves as a powerful tool to study the properties and structures of extremely proton-rich nuclei \cite{BLANK2008403}. Moreover, the orbital angular momentum of the emitted proton in the proton radioactivity carries the information of its wave function inside the nucleus \cite{Delion2006}. Up to now, a lot of empirical formulas or theoretical models have been developed to study the proton radioactivity, such as the single-folding model (SFM) \cite{Yi_Bin_2010, PhysRevC.72.051601}, the generalized liquid-drop model (GLDM) \cite{BAO201485}, the relativistic density functional theory (RDFT) \cite{FERREIRA2011508}, the effective interactions of density-dependent M3Y (DDM3Y) \cite{BHATTACHARYA2007263, Qian_2010}, the phenomenological unified fission model (PUFM) \cite{PhysRevC.71.014603, DONG2010198}, the method of coupled channel calculations (MCCC) \cite{PhysRevC.84.037301, PhysRevC.83.067302}, the distorted-wave Born approximation (DWBA) \cite{PhysRevC.56.1762}, the universal decay law of proton radioactivity (UDLP) \cite{PhysRevC.85.011303}, the Gamow-like model (GLM) \cite{Zdeb2016, Chen_2019}, the Coulomb and proximity potential model (CPPM) \cite{Deng2019} and so on. For more details about different theories of proton radioactivity, we refer the readers to Ref.\cite{Delion2006}.

The process of proton radioactivity can be treated as a quantum tunneling effect, whose penetration barrier can be obtained by the Wentzel-Kramers-Brillouin (WKB) approximation. The physical image of this process is similar to $\alpha$ decay. Recently, we used a two-potential approach (TPA) \cite{PhysRevLett.59.262} to systematically calculate $\alpha$ decay half-lives and proton radioactivity half-lives \cite{Sun_2017, PhysRevC.94.024338, PhysRevC.93.034316, PhysRevC.95.014319, PhysRevC.95.044303, PhysRevC.96.024318, PhysRevC.97.044322}. This approach was proposed by Gurvitz to deal with metastable states, and has been widely used to calculate $\alpha$ decay half-lives \cite{PhysRevLett.59.262}. For the study of $\alpha$ decay, we chose the nuclear potential of the cosh type proposed by Buck et al \cite{Buck1992}, while the Coulomb potential was taken as the potential of a uniformly charged sphere. For the study of proton radioactivity, both the emitted proton--daughter nucleus nuclear potential and Coulomb potential were calculated by Skyrme-Hartree-Fock (SHF) model \cite{doi:10.1063/1.2827286, PhysRevC.5.626}, which can successfully describe the structure of finite nuclei. For different purposes, more than 100 sets of Skyrme interactions have been constructed as of now. All these Skyrme interactions can reasonably describe the ground-state properties of nuclei over the periodic table and saturation properties of symmetric nuclear matter (SNM) \cite{Routray2012}. In this work, by employing the two-potential approach based on the Skyrme-Hartree-Fock model (TPA-SHF), we compare the performance of 115 sets of Skyrme interactions to study in the description of the spherical proton radioactivity half-lives of $53\leq Z \leq83$ nuclei. The Skyrme interaction that can best reproduce the experimental data are further used to predict the proton radioactivity half-lives of 7 unmeasured spherical nuclei in the same region.

This article is organized as follows. In Section \ref{section 2}, the theoretical framework for the proton radioactivity half-life is described in detail, including the two-potential approach and spherical Skyrme-Hartree-Fock model. In Section \ref{section 3}, the detailed calculations, a discussion and predictions are provided. A brief summary is given in Section \ref{section 4}.

\section{Theoretical framework}
\label{section 2}

The proton radioactivity half-life $T_\frac{1}{2}$ is inversely proportional to the proton radioactivity constant $\mathcal{\lambda}$ or decay width $\Gamma$ and can be written as 
\begin{equation}
\label{eq10}
T_\frac{1}{2}=\frac{ln2}{\lambda}=\frac{ln2\hbar}{\Gamma}.
\end{equation}
In the framework of the TPA \cite{PhysRevLett.59.262}, the $\Gamma$ depends on the normalized factor \emph{F} and the penetration probability of the emitted proton crossing the barrier $P$. It is given by
\begin{equation}
\label{eq11}
\Gamma=\frac{\hbar^2 F P}{4\mu},
\end{equation}
where the reduced mass $\mu=\frac{M_d M_p}{(M_d+M_p)}$, with $M_p$ and $M_d$ being the masses of the emitted proton and the daughter nuclei, and $\hbar$ is the reduced Planck constant. 

In the classical WKB approximation, the penetration probability $P$ is given by
\begin{equation}
\label{eq12}
P=\exp\! [-2 \int_{r_2}^{r_3} k(r)\, dr].
\end{equation}
Here the $r_2$, $r_3$ and the following $r_1$ represent the classical turning points, where the total emitted proton--daughter nucleus potential $V(r)$ is equal to the proton radioactivity energy $Q_p$. The wave number is $k(r)=\sqrt{\frac{2\mu}{\hbar^2}\left|Q_p-V(r)\right|}$. The total potential $V(r)$ can be written as

\begin{equation}
\label{eq1}
V(r)=V_N(r)+V_C(r)+V_l(r).
\end{equation}
where $V_N(r)$, $V_C(r)$ and $V_l(r)$ represent the nuclear, Coulomb, and centrifugal potentials, respectively.

In this work, the emitted proton--daughter nucleus nuclear potential $V_N(r)=U_q(\rho, \rho_p, \textbf{\emph{p}})$ is calculated by SHF with different Skyrme interactions, which are important in the finite nucleus calculations. The nuclear effective interactions with zero-range, momentum and density dependent forms is given by \cite{PhysRevC.5.626}
\begin{eqnarray}
\label{eq2}
V^{Skyrme}_{12}(\textbf{\emph{r}}_1, \textbf{\emph{r}}_2)=&t_0(1+x_0P_\sigma)\delta(\textbf{\emph{r}})\nonumber\\
&+ \frac{1}{2}t_1(1+x_1P_\sigma)[\textbf{\emph{P}}'^2\delta(\textbf{\emph{r}})+\delta(\textbf{\emph{r}})\textbf{\emph{P}}^2]
\nonumber\\
&+ t_2(1+x_2P_\sigma)\textbf{\emph{P}}'\cdot\delta(\textbf{\emph{r}})\textbf{\emph{P}}
\nonumber\\
&+ \frac{1}{6}t_3(1+x_3P_\sigma)[\rho(\textbf{\emph{R}})]^{\alpha}\delta(\textbf{\emph{r}})
\nonumber\\
&+ i {W_0} \bm{\sigma}\cdot[\textbf{\emph{P}}'\times\delta(\textbf{\emph{r}})\textbf{\emph{P}}], 
\end{eqnarray}
where $t_0$, $t_1$, $t_2$, $t_3$, $x_0$, $x_1$, $x_2$, $x_3$, $W_0$ and $\alpha$ are the Skyrme parameters. $\textbf{\emph{P}}'$ is the relative momentum operator acting on the left. $\textbf{\emph{P}}$ is its conjugate, which acts on the right. $P_\sigma$ represents the spin exchange operator. $\textbf{\emph{r}}_i$ (i=1, 2) is the coordinate vector of the $i$-$th$ nucleon. $\textbf{\emph{r}}=\textbf{\emph{r}}_1-\textbf{\emph{r}}_2$ and $\textbf{\emph{R}}=(\textbf{\emph{r}}_1+\textbf{\emph{r}}_2)/2$. In local density approximation, the single-nucleon potential from SHF model can be expressed as \cite{Zhang2018}

\begin{equation}
\label{eq3}
U_q(\rho, \rho_q, \textbf{\emph{p}})=a\textbf{\emph{p}}^2+b, 
\end{equation}
where $\textbf{\emph{p}}$ is the momentum of the nucleon. The coefficient $a$ and $b$ are given by

\begin{eqnarray}
\label{eq4}
a=&\frac{1}{8}[t_1(x_1+2)+t_2(x_2+2)]\rho\nonumber\\
&+ \frac{1}{8}[-t_1(2x_1+1)+t_2(2x_2+1)]\rho_q, 
\end{eqnarray}

\begin{eqnarray}
\label{eq5}
b=&\frac{1}{8}[t_1(x_1+2)+t_2(x_2+2)]\frac{k^5_{f, n}+k^5_{f, p}}{5\pi^2}\nonumber\\
&+\frac{1}{8}[t_2(2x_2+1)-t_1(2x_1+1)]\frac{k^5_{f, q}}{5\pi^2}
\nonumber\\
&+ \frac{1}{2}t_0(x_0+2)\rho-\frac{1}{2}t_0(2x_0+1)\rho_q
\nonumber\\
&+ \frac{1}{24}t_3(x_3+2)(\alpha+2)\rho^{(\alpha+1)}
\nonumber\\
&-\frac{1}{24}t_3(2x_3+1)\alpha\rho^{(\alpha-1)}(\rho^2_n+\rho^2_p)
\nonumber\\
&-\frac{1}{12}t_3(2x_3+1)\rho^\alpha\rho_q, 
\end{eqnarray}
where $\rho_q$ is the proton (neutron) density with $q=p$ $(n)$, $\rho=\rho_p+\rho_n$ is the total nucleon density, and $k_{f, q}=(3\pi\rho_q)^{1/3}$ is the Fermi momentum of the nucleon. The total energy $E$ of a nucleon in a nuclear medium can be written as
\begin{eqnarray}
\label{eq6}
E=&U_q(\rho, \rho_q, \textbf{\emph{p}})+\frac{\textbf{\emph{p}}^2}{2m}+V_C(r)\nonumber\\
&=\frac{\textbf{\emph{p}}^2}{2m^{*}}+b+V_C(r), 
\end{eqnarray}
where $m^{*}$ stands for effective mass defined by 
\begin{equation}
\label{eq7}
\frac{1}{2m^{*}}=\frac{1}{2m}+a.
\end{equation}
Under the premise that the total energy of the proton remains constant during proton radioactivity, the momentum of the proton $\left|\textbf{\emph{p}}\right|$ can be derived from isospin asymmetry, the total energy and the nucleon density. It can be given by

\begin{equation}
\left|\textbf{\emph{p}}\right|=\sqrt{2m^{*}(E-b-V_C(r))}, 
\end{equation}
and the potential energy $U_q(\rho, \rho_p, \textbf{\emph{p}})=E-\frac{\textbf{\emph{p}}^2}{2m}-V_C(r)$.

The Coulomb potential $V_C(r)$ comprises the direct $V^{d}_C(r)$ and exchange part $V^{ex}_C(r)$ \cite{Routray2012}, $V_C(r)=V^{d}_C(r)+V^{ex}_C(r)$, given as
\begin{equation}
V^{d}_C(r)=4\pi e^2 [\frac{1}{r} \int_{0}^{r} r'^2\rho_p(r')\, dr']+[ \int_{r}^{\infty} r'\rho_p(r')\, dr'],
\end{equation}

\begin{equation}
V^{ex}_C(r)=-e^2[\frac{3}{\pi}\rho_p(r)]^\frac{1}{3}.
\end{equation}

For centrifugal potential $V_{l}(r)$, $l(l + 1) \rightarrow (l + 1/2)^2$ is an essential correction \cite{PhysRevC.82.059902}. In this work, the centrifugal potential $V_{l}(r)$ is chosen as the Langer-modified form. It can be written as
\begin{equation}
\label{eq9}
V_{l}(r)=\frac{\hbar^2(l+\frac{1}{2})^2}{2{\mu}r^2}, 
\end{equation}
where $l$ is the orbital angular momentum taken away by the emitted proton \cite{doi:10.1063/1.531270}. It can be obtained by the parity and angular momentum conservation laws.

The normalized factor $F$, determining the assault frequency, can be obtained by the integration over the internal region \cite{PhysRevLett.59.262}. It is expressed as
\begin{equation}
\label{eq3}
F\! \int_{r_1}^{r_2}\frac{1}{2k(r)}\, dr=1.
\end{equation}

\section{Results and discussion}
\label{section 3}

The aim of the present work is to perform a comparative study of different Skyrme interactions when they are used to calculate the proton radioactivity half-lives of spherical proton emitters and determine which Skyrme interaction can give the lowest r.m.s. deviation in the description of the experimental data. The experimental proton radioactivity half-lives, spin/parity, $Q_p$ and its uncertainties are taken from the latest evaluated nuclear properties table NUBASE2016 \cite{1674-1137-41-3-030001} and the latest evaluated atomic mass table AME2016 \cite{1674-1137-41-3-030002,1674-1137-41-3-030003} except for those of $^{144}$Tm, $^{150}$Lu, $^{151}$Lu, $^{159}$Re, $^{159}$Re$^m$, and $^{164}$Ir, which are taken from Ref. \cite{BLANK2008403}. First, the proton radioactivity half-lives of 32 parent nuclei with $53\leq Z \leq83$ are studied with 115 sets of Skyrme parameters. For different Skyrme interactions, the standard deviation $\sigma=\sqrt{\sum ({\rm{lg}{T^{\rm{exp}}_{1/2}}(s)}-{\rm{lg}{T^{\rm{cal}}_{1/2}}(s)})^2/n}$ between proton radioactivity half-lives calculated by TPA-SHF and experimental data are listed in Table \ref{tab1} in ascending order. As seen from this table, the standard deviations corresponding to various versions of Skyrme interactions are different. They range from 0.3869 to 0.5464, and the proton radioactivity half-lives calculated by the Skyrme interaction of SLy7 are closest to the experimental data. Meanwhile, for most of the 115 Skyrme interactions, the obtained standard deviations are less than 0.5, and only 6 Skyrme interactions give larger the standard deviations lying in the range of $0.5-0.6$.

\begin{table*}[hbt!]
\caption{Standard deviation $\sigma$ between proton radioactivity half-lives calculated by TPA-SHF with 115 Skyrme interactions and experimental data.}
\label{tab1}
\renewcommand\arraystretch{1.2}
\setlength{\tabcolsep}{4.0mm}
\begin{footnotesize}
\begin{threeparttable}
\begin{tabular}{|ccc|ccc|ccc|}
\hline
 Skyrme set & $\sigma$ & Ref. & Skyrme set & $\sigma$ & Ref. & Skyrme set & $\sigma$ & Ref. \\
 \hline
SLy7	&	0.3869	&	\cite{CHABANAT1998231}	&	BSk12&	0.4158	&	\cite{GORIELY2006279}	&	SKM 	&	0.4328	&	\cite{KRIVINE1980155}	\\
SLy3	&	0.3929	&	\cite{PhysRevC.85.024305}	&	BSk13&	0.416	&	\cite{GORIELY2006279}	&	SIV 	&	0.4351	&	\cite{BEINER197529}	\\
SV 	&	0.3955	&	\cite{PhysRevC.33.335}	&	SKM*	&	0.4166	&	\cite{BARTEL198279}	&	SKP 	&	0.4352	&	\cite{PhysRevC.58.2099}	\\
SLy230b 	&	0.3956	&	\cite{CHABANAT1997710}	&	SKT3*&	0.4167	&	\cite{TONDEUR1984297}	&	SKT9	&	0.4352	&	\cite{TONDEUR1984297}	\\
SKT5	&	0.3969	&	\cite{TONDEUR1984297}	&	SKT2	&	0.417	&	\cite{TONDEUR1984297}	&	SkSC4&	0.4356	&	\cite{ABOUSSIR1992155}	\\
BSk9	&	0.4001	&	\cite{PhysRevC.70.044309}	&	SKT8	&	0.4172	&	\cite{TONDEUR1984297}	&	SkSC11  	&	0.4356	&	\cite{PhysRevC.52.2254}	\\
RATP	&	0.4003	&	\cite{m1}	&	 MSk2&	0.4174	&	\cite{PhysRevC.62.024308}	&	BSk4	&	0.4359	&	\cite{PhysRevC.68.054325}	\\
SGI 	&	0.4007	&	\cite{VANGIAI1981379}	&	BSk14&	0.4174	&	\cite{PhysRevC.75.064312}	&	SkSC15  	&	0.4366	&	\cite{PEARSON2000163}	\\
Skz4	&	0.4013	&	\cite{PhysRevC.66.014303}	&	 MSk5&	0.42	&	\cite{PhysRevC.62.024308}	&	Zsigma-fit	&	0.4368	&	\cite{PhysRevC.33.335}	\\
Skz1	&	0.4017	&	\cite{PhysRevC.66.014303}	&	 v0.80  	&	0.4202	&	\cite{PhysRevC.64.027301}	&	 v0.70  	&	0.438	&	\cite{PhysRevC.64.027301}	\\
KDE0	&	0.4025	&	\cite{PhysRevC.72.014310}	&	 BSk2p  	&	0.4203	&	\cite{PhysRevC.66.024326}	&	SkSC6&	0.4385	&	\cite{PhysRevC.50.460}	\\
Skz3	&	0.4028	&	\cite{PhysRevC.66.014303}	&	 Dutta  	&	0.4214	&	\cite{DUTTA1986374}	&	 v0.90  	&	0.4386	&	\cite{PhysRevC.64.027301}	\\
SLy0	&	0.4037	&	\cite{PhysRevC.85.024305}	&	SkSC14  	&	0.4216	&	\cite{PEARSON2000163}	&	 MSk9&	0.4387	&	\cite{GORIELY2001349}	\\
MSk7	&	0.4042	&	\cite{GORIELY2001311}	&	Esigma-fit	&	0.4217	&	\cite{PhysRevC.33.335}	&	SIII	&	0.4388	&	\cite{BEINER197529}	\\
SKT4	&	0.4049	&	\cite{TONDEUR1984297}	&	BSk1	&	0.4221	&	\cite{SAMYN2002142}	&	BSk7	&	0.4401	&	\cite{PhysRevC.68.054325}	\\
SGII	&	0.4049	&	\cite{VANGIAI1981379}	&	SkSC2&	0.4223	&	\cite{PEARSON19911}	&	SVI 	&	0.4408	&	\cite{PhysRevC.21.2076}	\\
SLy2	&	0.4054	&	\cite{PhysRevC.85.024305}	&	SkSC3&	0.4224	&	\cite{PEARSON19911}	&	 MSk6&	0.4417	&	\cite{PhysRevC.62.024308}	\\
 MSk3&	0.4059	&	\cite{PhysRevC.62.024308}	&	SKa 	&	0.4228	&	\cite{KOHLER1976301}	&	SKO*	&	0.4423	&	\cite{PhysRevC.60.014316}	\\
SLy5	&	0.406	&	\cite{CHABANAT1998231}	&	SkI2	&	0.4228	&	\cite{REINHARD1995467}	&	 SVII&	0.4431	&	\cite{PhysRevC.21.2076}	\\
 v0.75  	&	0.4062	&	\cite{PhysRevC.64.027301}	&	BSk3	&	0.4229	&	\cite{PhysRevC.68.054325}	&	Rsigma-fit	&	0.4439	&	\cite{PhysRevC.33.335}	\\
SkMP	&	0.4071	&	\cite{PhysRevC.40.2834}	&	SkSC4o  	&	0.4237	&	\cite{PEARSON2000163}	&	 SK255  	&	0.4474	&	\cite{PhysRevC.68.031304}	\\
SkSC5&	0.4071	&	\cite{PhysRevC.50.460}	&	SkI3	&	0.4239	&	\cite{REINHARD1995467}	&	Gsigma-fit	&	0.4479	&	\cite{PhysRevC.33.335}	\\
SkI5	&	0.4078	&	\cite{REINHARD1995467}	&	BSk11&	0.424	&	\cite{GORIELY2006279}	&	BSk8	&	0.4528	&	\cite{PhysRevC.70.044309}	\\
SLy1	&	0.4084	&	\cite{PhysRevC.85.024305}	&	 v1.00  	&	0.4243	&	\cite{PhysRevC.64.027301}	&	SIII*&	0.4528	&	\cite{PhysRevC.21.2076}	\\
 BSk17  	&	0.4087	&	\cite{PhysRevLett.102.152503} 	&	 MSk8&	0.4248	&	\cite{GORIELY2001349}	&	SKXm	&	0.4534	&	\cite{PhysRevC.58.2099}	\\
MSk1	&	0.4087	&	\cite{PhysRevC.62.024308}	&	SkSC10  	&	0.4263	&	\cite{PhysRevC.50.460}	&	BSk10&	0.4556	&	\cite{GORIELY2006279}	\\
SLy10&	0.4092	&	\cite{PhysRevC.85.024305}	&	SkSC1&	0.4263	&	\cite{PEARSON19911}	&	BSk15&	0.4611	&	\cite{PhysRevC.77.031301}	\\
SKT6	&	0.4094	&	\cite{TONDEUR1984297}	&	BSk2	&	0.4267	&	\cite{PhysRevC.66.024326}	&	SKXce&	0.4704	&	\cite{PhysRevC.58.2099}	\\
SKb 	&	0.4104	&	\cite{VANGIAI1981379}	&	SKT1*&	0.4278	&	\cite{TONDEUR1984297}	&	 SKY T  	&	0.4714	&	\cite{KO1974269}	\\
Skz2	&	0.4105	&	\cite{PhysRevC.66.014303}	&	SLy4	&	0.4288	&	\cite{CHABANAT1998231}	&	SKX 	&	0.4828	&	\cite{PhysRevC.58.2099}	\\
BSk6	&	0.4112	&	\cite{PhysRevC.68.054325}	&	SLy8	&	0.429	&	\cite{PhysRevC.85.024305}	&	 LNS	&	0.4982	&	\cite{PhysRevC.73.014313}	\\
 v1.10  	&	0.4123	&	\cite{PhysRevC.64.027301}	&	SKT7	&	0.4297	&	\cite{TONDEUR1984297}	&	Z-fit&	0.5071	&	\cite{PhysRevC.33.335}	\\
SLy230a 	&	0.4128	&	\cite{CHABANAT1998231}	&	BSk16&	0.4298	&	 \cite{CHAMEL200872}	&	SG0I	&	0.5083	&	\cite{Treiner_1976}	\\
SLy9	&	0.4132	&	\cite{PhysRevC.85.024305}	&	SII 	&	0.4299	&	\cite{PhysRevC.5.626}	&	Skyrme 1p	&	0.5291	&	\cite{PETHICK1995675}	\\
 MSk4&	0.4139	&	\cite{PhysRevC.62.024308}	&	 SK272  	&	0.4306	&	\cite{PhysRevC.68.031304}	&	SI 	&	0.5405	&	\cite{PhysRevC.33.335}	\\
SKT1	&	0.4139	&	\cite{TONDEUR1984297}	&	BSk5	&	0.431	&	\cite{PhysRevC.68.054325}	&	E-fit&	0.5424	&	\cite{PhysRevC.33.335}	\\
SkI6	&	0.4152	&	\cite{PhysRevC.53.740}	&	Zsigma*-fi	&	0.4314	&	\cite{PhysRevC.33.335}	&	MSkA	&	0.5464	&	\cite{PhysRevLett.74.3744}	\\
KDE 	&	0.4158	&	\cite{PhysRevC.72.014310}	&	SKT3	&	0.4321	&	\cite{TONDEUR1984297}	&		&		&		\\
SLy6	&	0.4158	&	\cite{CHABANAT1998231}	&	 v1.05  	&	0.4324	&	\cite{PhysRevC.64.027301}	&		&		&		\\

\hline
\end{tabular}
\end{threeparttable}
\end{footnotesize}
\end{table*}

In this work, for the same nucleus, the differences in the theoretical proton radioactivity half-lives come from the different nuclear potentials $V_N(r)$ and Coulomb potentials $V_C(r)$ obtained by different Skyrme interactions. To compare experimental data with the theoretical proton radioactivity half-lives of the same nucleus calculated by different Skyrme interactions, we show the results in Table \ref{tab2}. In this table, the first three columns present the parent nuclei, the $l$ value, and the proton radioactivity energy $Q_p$. The next column presents the logarithmic form of the experimental proton radioactivity half-lives. The last eight columns presents the logarithmic form of the theoretical proton radioactivity half-life calculated by TPA-SHF within the Skyrme interactions of SLy7, SLy3, SV, SLy230b, Skyrme 1p, SI, E-fit and MSkA. The corresponding errors in the calculated half-lives and experimental errors in $Q_p$ values are inside parentheses. While the first four Skyrme interactions of SLy7, SLy3, SV and SLy230b correspond to the four minimum $\sigma$, the last four Skyrme interactions of Skyrme 1p, SI, E-fit and MSkA correspond to the four maximum $\sigma$.

\begin{table*}[hbt!]
\caption{Logarithmic forms of theoretical proton radioactivity half-lives corresponding to different Skyrme interactions compared with experimental data.}
\label{tab2}
\renewcommand\arraystretch{1.4}
\setlength{\tabcolsep}{0.4mm}
\begin{footnotesize}
\begin{threeparttable}
\begin{tabular}{cccccccccccc}
\hline
Parent nuclei & $l$ & ${Q_p}$& Measured &SLy7& SLy3 & SV & SLy230b &Skyrme 1p & SI & E-fit & MSkA \\
$^{A}Z$ & &KeV& $\rm{lg}{T_{1/2}}$(s)& $\rm{lg}{T_{1/2}}$(s)& $\rm{lg}{T_{1/2}}$(s)& $\rm{lg}{T_{1/2}}$(s)& $\rm{lg}{T_{1/2}}$(s)& $\rm{lg}{T_{1/2}}$(s)& $\rm{lg}{T_{1/2}}$(s)& $\rm{lg}{T_{1/2}}$(s)& $\rm{lg}{T_{1/2}}$(s) \\
 \hline
$^{144}$Tm&5&1725(16)&-5.569&-5.24(0.16)&-5.24(0.1)&-5.23(0.14)&-5.21(0.13)&-4.81(0.27)&-4.75(0.33)&-5.04(0.12)&-4.92(0.11)\\
$^{145}$Tm &5&1736(7)&-5.499&-5.34(0.05)&-5.32(0.16)&-5.31(0.05)&-5.36(0.04)&-5.08(0.05)&-5.05(0.1)&-5.14(0.05)&-5.02(0.04)\\
$^{146}$Tm &0&896(6)&-0.81&-0.6(0.1)&-0.56(0.17)&-0.54(0.17)&-0.62(0.14)&-0.33(0.11)&-0.38(0.1)&-0.38(0.15)&-0.27(0.07)\\
$^{146}$Tm$^m$&5&1199.3(1)&-1.125&-0.67(0.05)&-0.7(0.09)&-0.69(0.08)&-0.59(0.09)&-0.44(0.03)&-0.41(0.07)&-0.48(0.02)&-0.32(0.07)\\
$^{147}$Tm &5&1059(3)&0.573&1.07(0.15)&1.03(0.08)&1.06(0.13)&1.09(0.09)&1.29(0.06)&1.28(0.04)&1.24(0.06)&1.36(0.07)\\
$^{147}$Tm$^m$&2&1118.5(3.9)&-3.444&-3.03(0.08)&-3.03(0.06)&-3(0.05)&-3.05(0.06)&-2.74(0.11)&-2.78(0.04)&-2.84(0.13)&-2.75(0.07)\\
$^{150}$Lu$^m$&5&1283(3)&-1.18&-1.07(0.03)&-1.08(0.03)&-1.08(0.04)&-1.09(0.03)&-0.66(0.22)&-0.8(0.08)&-0.89(0.05)&-0.79(0.04)\\
$^{150}$Lu&2&1294.7(6)&-4.398&-4.37(0.16)&-4.42(0.14)&-4.42(0.06)&-4.39(0.05)&-4.13(0.09)&-4.18(0.07)&-4.16(0.12)&-4.16(0.44)\\
$^{151}$Lu$^m$&5&1253(3)&-0.896&-0.77(0.05)&-0.78(0.09)&-0.76(0.03)&-0.76(0.06)&-0.52(0.02)&-0.28(0.28)&-0.56(0.04)&-0.49(0.05)\\
$^{151}$Lu&2&1293.6(4)&-4.783&-4.42(0.14)&-4.42(0.04)&-4.42(0.05)&-4.43(0.03)&-4.09(0.11)&-4.15(0.04)&-4.13(0.02)&-4.15(0.07)\\
$^{155}$Ta &5&1453(15)&-2.495&-2.25(0.13)&-2.22(0.14)&-2.23(0.2)&-2.28(0.15)&-1.99(0.13)&-1.98(0.14)&-2.07(0.17)&-1.99(0.15)\\
$^{156}$Ta &2&1020(4)&-0.828&-0.37(0.17)&-0.38(0.07)&-0.38(0.13)&-0.41(0.08)&-0.14(0.31)&-0.06(0.16)&-0.17(0.07)&-0.05(0.07)\\
$^{156}$Ta$^m$&5&1115.2(8)&0.924&1.48(0.14)&1.49(0.14)&1.55(0.14)&1.42(0.13)&1.69(0.12)&1.69(0.09)&2.51(0.89)&1.77(0.12)\\
$^{157}$Ta &0&935(10)&-0.529&0.12(0.21)&0.12(0.18)&0.14(0.26)&0.11(0.17)&0.37(0.22)&0.58(0.35)&0.31(0.23)&0.44(0.2)\\
$^{159}$Re&5&1816(20)&-4.678&-4.59(0.13)&-4.49(0.29)&-4.57(0.18)&-4.65(0.14)&-4.33(0.07)&-4.39(0.13)&-4.43(0.14)&-4.31(0.14)\\
$^{159}$Re$^m$&5&1831(20)&-4.695&-4.68(0.17)&-4.73(0.15)&-4.66(0.16)&-4.39(0.5)&-4.48(0.15)&-4.48(0.13)&-4.44(0.22)&-4.41(0.15)\\
$^{160}$Re &0&1267(7)&-3.164&-3.76(0.13)&-3.84(0.55)&-3.75(0.13)&-3.81(0.11)&-3.49(0.14)&-3.59(0.09)&-3.6(0.09)&-3.39(0.23)\\
$^{161}$Re &0&1197(5)&-3.357&-2.95(0.14)&-3(0.06)&-3(0.16)&-3.03(0.08)&-2.78(0.07)&-2.78(0.05)&-2.72(0.16)&-2.72(0.05)\\
$^{161}$Re$^m$&5&1323.3(7)&-0.68&-0.51(0.08)&-0.44(0.1)&-0.5(0.09)&-0.54(0.14)&-0.28(0.08)&-0.27(0.08)&-0.32(0.07)&-0.16(0.08)\\
$^{164}$Ir&5&1844(9)&-3.947&-4.41(0.08)&-4.34(0.15)&-4.39(0.05)&-4.43(0.04)&-4.1(0.09)&-4.15(0.24)&-4.22(0.09)&-4.11(0.06)\\
$^{165}$Ir$^m$&5&1717.5(7)&-3.43&-3.55(0.1)&-3.57(0.11)&-3.54(0.06)&-3.58(0.21)&-3.2(0.15)&-3.32(0.16)&-3.35(0.33)&-3.19(0.05)\\
$^{166}$Ir &2&1152(8)&-0.842&-0.98(0.13)&-1.03(0.12)&-1(0.12)&-1.03(0.13)&-0.75(0.1)&-0.64(0.25)&-0.75(0.12)&-0.71(0.14)\\
$^{166}$Ir$^m$&5&1324.1(8)&-0.091&0.05(0.1)&0.02(0.15)&0.2(0.29)&-0.03(0.11)&0.22(0.11)&0.22(0.11)&0.21(0.12)&0.33(0.13)\\
$^{167}$Ir &0&1070(4)&-1.128&-0.7(0.06)&-0.71(0.07)&-0.63(0.08)&-0.69(0.1)&-0.45(0.13)&-0.46(0.08)&-0.51(0.05)&-0.39(0.35)\\
$^{167}$Ir$^m$&5&1245.5(7)&0.778&0.86(0.08)&0.86(0.09)&0.9(0.08)&0.87(0.09)&1.1(0.11)&1.11(0.09)&1.08(0.08)&1.2(0.17)\\
$^{170}$Au &2&1472(12)&-3.487&-3.93(0.21)&-4.04(0.14)&-3.98(0.13)&-3.9(0.26)&-3.68(0.12)&-3.78(0.15)&-3.8(0.16)&-3.74(0.14)\\
$^{170}$Au$^m$&5&1753.5(6)&-2.975&-3.13(0.28)&-3.16(0.3)&-3.25(0.18)&-3.34(0.13)&-3.16(0.05)&-3.16(0.05)&-3.21(0.06)&-3.1(0.12)\\
$^{171}$Au &0&1448(10)&-4.652&-4.61(0.13)&-4.64(0.14)&-4.57(0.36)&-4.65(0.1)&-4.38(0.11)&-4.37(0.1)&-4.42(0.12)&-4.3(0.09)\\
$^{171}$Au$^m$&5&1703(4)&-2.587&-3(0.07)&-3.02(0.21)&-2.9(0.07)&-3.03(0.06)&-2.79(0.77)&-2.77(0.32)&-2.84(0.25)&-2.74(0.12)\\
$^{176}$Tl &0&1265(18)&-2.208&-2.06(0.21)&-2.1(0.26)&-1.98(0.24)&-2.07(0.25)&-1.82(0.23)&-1.77(0.34)&-1.87(0.21)&-1.74(0.28)\\
$^{177}$Tl &0&1155(19)&-1.178&-0.66(0.28)&-0.66(0.28)&-0.61(0.31)&-0.67(0.27)&-0.23(0.44)&-0.41(0.8)&-0.46(0.27)&-0.38(0.59)\\
$^{177}$Tl$^m$&5&1968(8)&-3.459&-4.52(0.04)&-4.52(0.07)&-4.46(0.04)&-4.55(0.08)&-4.27(0.13)&-4.26(0.05)&-4.31(0.2)&-4.19(0.09)\\

\hline
\end{tabular}
\end{threeparttable}
\end{footnotesize}
\end{table*}

\begin{figure*}[h]
\centerline{\includegraphics[width=10cm]{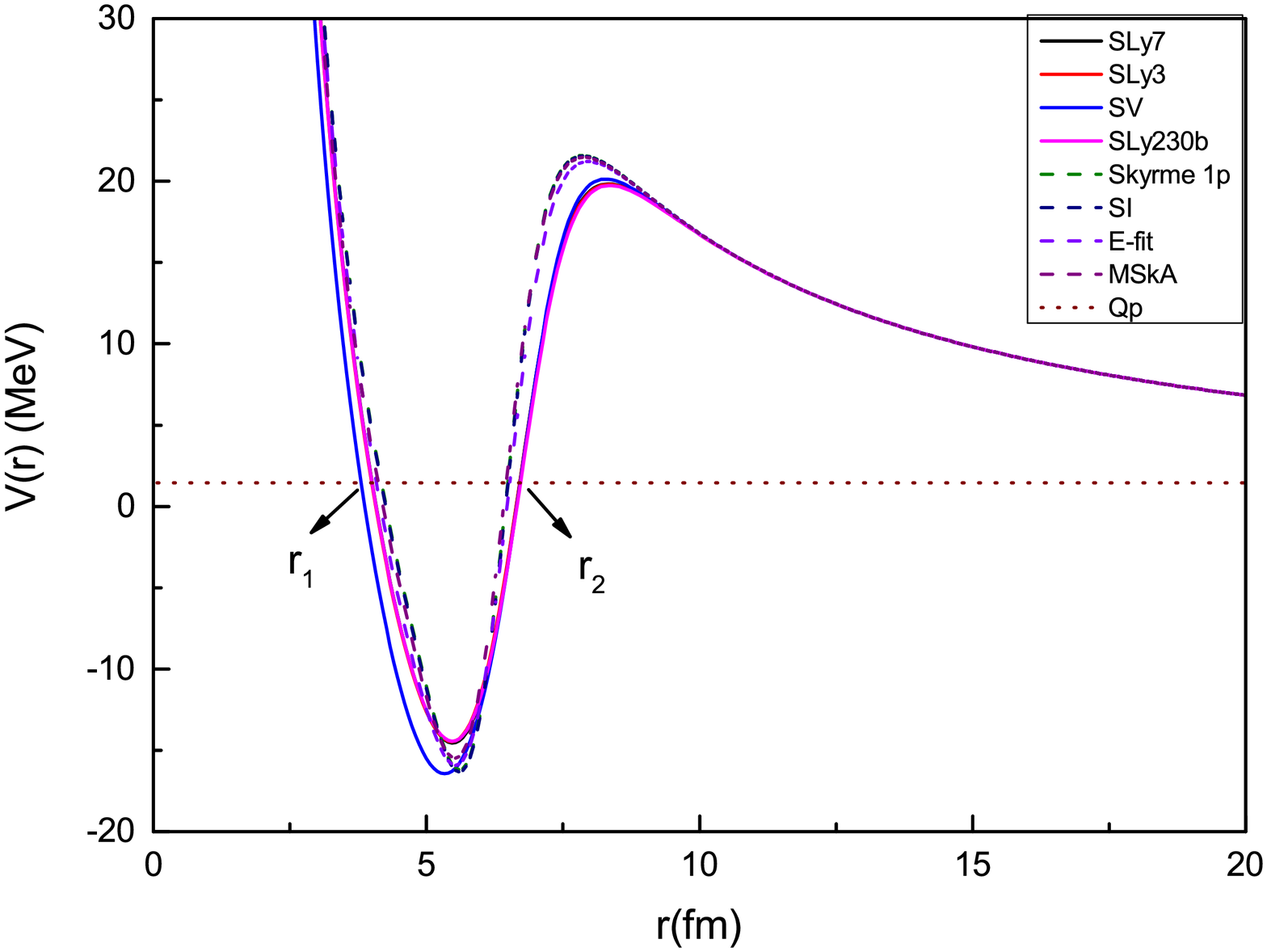}}
\caption{(Color online) Total potential $V(r)$ distribution corresponding to the eight versions of Skyrme interactions for $^{155}$Ta.}
\label{fig1}
\end{figure*}

From Table \ref{tab2}, we can clearly see that most of the theoretical proton radioactivity half-lives calculated by TPA-SHF can reproduce the experimental data very well. Furthermore, the calculated results of the first four Skyrme interactions are significantly different from those of the last four Skyrme interactions. For the same parent nuclei, the theoretical proton radioactivity half-lives calculated by the last four Skyrme interactions are greater than those calculated by the first four Skyrme interactions. Based on our comparative analysis of the theoretical proton radioactivity half-lives obtained using different Skyrme interactions, we can study what particular feature of a given potential impacts these differences between theoretical calculations and the differences between theory and experiment. It can be found from Sect. \ref{section 2} that the proton radioactivity half-life can be obtained by the normalized factor \emph{F} and the penetration probability $P$, which depends on proton radioactivity energy $Q_p$ and the total emitted proton-daughter nucleus interaction potential $V(r)$. A higher barrier will result in a longer theoretical proton radioactivity half-life. The reason why the theoretical proton radioactivity half-lives calculated by the last four Skyrme interactions cannot well reproduce the experimental data may be that the nuclear potential value obtained by these Skyrme interactions is greater than the true nuclear potential. To verify this conclusion, taking $^{155}$Ta as an example, we plot the total potential $V(r)$ distribution from the eight Skyrme interactions in Figure \ref{tab1}. The results from SLy7, SLy3, SV and SLy230b are shown as solid lines with different colors, while those from Skyrme 1p, SI, E-fit and MSkA dashed lines with different colors. To identify the position of the turning points $r_1$ and $r_2$, the proton radioactivity energy $Q_p$ is indicated in Fig.\ref{fig1} by the dotted line. 

Figure \ref{tab1} shows that when the range of $r$ is from 0$fm$ to $r_1$ or greater than 10$fm$, the curves of the various versions of $V(r)$ almost coincide. Moreover, the dashed and solid lines are clearly divided into two groups within the range of $r_2<r<10fm$.The $V(r)$ corresponding to the dashed line is greater than the solid line. This rule confirms that the too large $V(r)$ after $r>r_2$ leads to overestimate of the half-lives of proton radioactivity by some Skrme interactions. Among all the potentials presented in Fig. \ref{fig1}, it is the one obtained from SLy7 that provides the best description of the proton decay half-lives from Table \ref{tab1}. 

\begin{figure*}[h]
\centerline{\includegraphics[width=10cm]{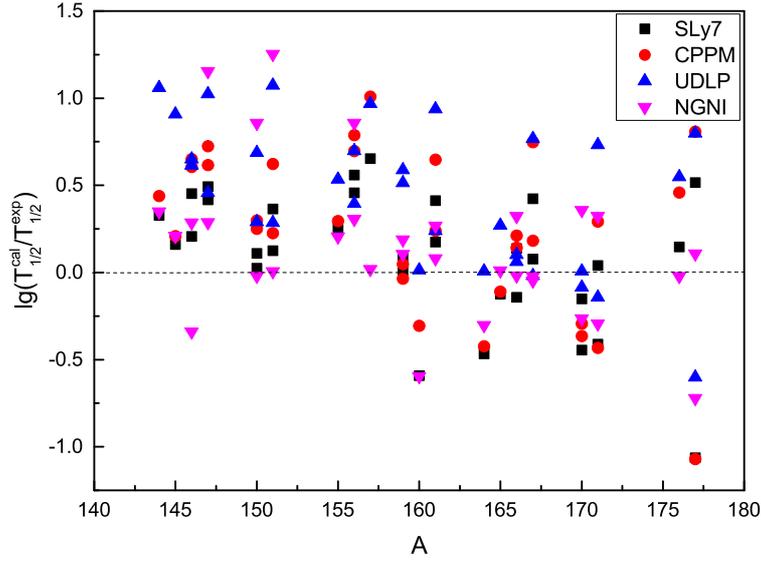}}
\caption{(Color online) Deviations between $\rm{lg}{T^{cal}_{1/2}}$ and $\rm{lg}{T^{exp}_{1/2}}$. The 4 different colors and shapes correspond to the calculations obtained by 4 different models.}
\label{fig2}
\end{figure*}

In 2012, Qi \emph{et al} extended the universal decay law (UDL) \cite{PRL2012} for the calculation of the proton radioactivity half-life (UDLP)\cite{PhysRevC.85.011303}. In 2019, Deng \emph{et al} conducted a comparative study of the 28 proximity potential formalisms for the calculation of the proton radioactivity half-lives of spherical proton emitters. They found that the proximity potential Guo2013 formalism (CPPM-Guo2013) gives the lowest r.m.s. for spherical proton emitters \cite{PhysRevC.85.011303}. Recently, Chen \emph{et al} proposed a two-parameter formula of the new Geiger-Nuttall law (NGNL) for proton radioactivity, which can well reproduce the experimental data of the half-life \cite{Chen20192}. For comparison, we calculated the proton radioactivity half-lives with these models and list the logarithmic form of theoretical calculations together with the experimental data in Table \ref{tab5}. In this table, the first four columns are the same as in Table \ref{tab2}, and the next three columns present the logarithmic form of the theoretical proton radioactivity half-lives calculated by CPPM-Guo2013, UDLP and NGNL, denoted CPPM, UDLP and NGNL, respectively. The corresponding errors in the calculated half-lives and experimental errors in $Q_p$ values are inside parentheses. To intuitively compare these models with ours, 
the logarithmic deviations between experimental data of proton radioactivity half-lives and theoretically calculated ones are shown in Figure \ref{fig2}. In this figure, the squares, circles, triangles and inverted triangles correspond to the results obtained by SLy7 (TPA-SHF-SLy7), UDLP, CPPM-Guo2013, and NGNL, denoted SLy7, UDLP, CPPM, and NGNL, respectively. It can be seen from Figure \ref{fig2} that our calculations can better reproduce the experimental data than other models. Moreover, we also calculated the standard deviation $\sigma$ between the logarithmic theoretical values calculated by these models and experimental data. The results for $\rm{T}_{1/2}^{\rm{SLy7}}$ , $\rm{T}_{1/2}^{\rm{CPPM}}$, $\rm{T}_{1/2}^{\rm{UDLP}}$ and $\rm{T}_{1/2}^{\rm{NGNL}}$ are listed in Table \ref{tab3}. These results show that the TPA-SHF-SLy7 method is better than the other models in calculating the spherical proton radioactivity half-lives. 

\begin{table*}[hbt!]
\caption{Logarithmic forms of theoretical proton radioactivity half-lives calculated by CPPM-Guo2013, UDLP and NGNL.}
\label{tab5}
\renewcommand\arraystretch{1.4}
\setlength{\tabcolsep}{6mm}
\begin{footnotesize}
\begin{threeparttable}
\begin{tabular}{ccccccc}
\hline
Parent nuclei & $l$ & ${Q_p}$ & Measured &CPPM& UDLP & NGNL \\
$^{A}Z$ & &KeV& $\rm{lg}{T_{1/2}}$(s)& $\rm{lg}{T_{1/2}}$(s) &$\rm{lg}{T_{1/2}}$(s) & $\rm{lg}{T_{1/2}}$(s) \\
 \hline
$^{144}$Tm&5&1725(16)&-5.569&-5.13(0.11)&-4.51(0.09)&-5.22(0.1)\\
$^{145}$Tm &5&1736(7)&-5.499&-5.29(0.05)&-4.59(0.04)&-5.29(0.04)\\
$^{146}$Tm &0&896(6)&-0.81&-0.16(0.11)&-0.16(0.09)&-1.15(0.09)\\
$^{146}$Tm$^m$&5&1199.3(1)&-1.125&-0.52(0.01)&-0.51(0.01)&-0.84(0.01)\\
$^{147}$Tm &5&1059(3)&0.573&1.19(0.04)&1.03(0.04)&0.86(0.04)\\
$^{147}$Tm$^m$&2&1118.5(3.9)&-3.444&-2.72(0.05)&-2.42(0.04)&-2.29(0.04)\\
$^{150}$Lu$^m$&5&1283(3)&-1.18&-0.93(0.03)&-0.89(0.03)&-1.2(0.03)\\
$^{150}$Lu&2&1294.7(6)&-4.398&-4.1(0.06)&-3.71(0.06)&-3.54(0.06)\\
$^{151}$Lu$^m$&5&1253(3)&-0.896&-0.67(0.03)&-0.61(0.03)&-0.89(0.03)\\
$^{151}$Lu&2&1293.6(4)&-4.783&-4.16(0.04)&-3.71(0.04)&-3.53(0.04)\\
$^{155}$Ta &5&1453(15)&-2.495&-2.2(0.14)&-1.96(0.12)&-2.29(0.13)\\
$^{156}$Ta &2&1020(4)&-0.828&-0.04(0.06)&-0.13(0.05)&0.03(0.05)\\
$^{156}$Ta$^m$&5&1115.2(8)&0.924&1.62(0.11)&1.32(0.1)&1.23(0.1)\\
$^{157}$Ta &0&935(10)&-0.529&0.48(0.18)&0.44(0.16)&-0.51(0.14)\\
$^{159}$Re&5&1816(20)&-4.678&-4.63(0.14)&-4.09(0.12)&-4.49(0.13)\\
$^{159}$Re$^m$&5&1831(20)&-4.695&-4.73(0.14)&-4.18(0.12)&-4.59(0.12)\\
$^{160}$Re &0&1267(7)&-3.164&-3.47(0.08)&-3.15(0.07)&-3.76(0.07)\\
$^{161}$Re &0&1197(5)&-3.357&-2.71(0.06)&-2.42(0.06)&-3.09(0.05)\\
$^{161}$Re$^m$&5&1323.3(7)&-0.68&-0.44(0.08)&-0.44(0.07)&-0.6(0.07)\\
$^{164}$Ir&5&1844(9)&-3.947&-4.37(0.06)&-3.94(0.05)&-4.25(0.06)\\
$^{165}$Ir$^m$&5&1717.5(7)&-3.43&-3.54(0.05)&-3.16(0.05)&-3.42(0.05)\\
$^{166}$Ir &2&1152(8)&-0.842&-0.7(0.11)&-0.74(0.1)&-0.52(0.09)\\
$^{166}$Ir$^m$&5&1324.1(8)&-0.091&0.12(0.09)&-0.03(0.08)&-0.11(0.08)\\
$^{167}$Ir &0&1070(4)&-1.128&-0.38(0.06)&-0.36(0.05)&-1.15(0.05)\\
$^{167}$Ir$^m$&5&1245.5(7)&0.778&0.96(0.09)&0.76(0.07)&0.73(0.08)\\
$^{170}$Au &2&1472(12)&-3.487&-3.78(0.12)&-3.48(0.1)&-3.13(0.1)\\
$^{170}$Au$^m$&5&1753.5(6)&-2.975&-3.34(0.05)&-3.06(0.04)&-3.24(0.04)\\
$^{171}$Au &0&1448(10)&-4.652&-4.36(0.1)&-3.92(0.09)&-4.33(0.08)\\
$^{171}$Au$^m$&5&1703(4)&-2.587&-3.02(0.03)&-2.73(0.03)&-2.88(0.03)\\
$^{176}$Tl &0&1265(18)&-2.208&-1.75(0.22)&-1.66(0.2)&-2.23(0.18)\\
$^{177}$Tl &0&1155(19)&-1.178&-0.37(0.27)&-0.38(0.24)&-1.07(0.22)\\
$^{177}$Tl$^m$&5&1968(8)&-3.459&-4.53(0.05)&-4.06(0.05)&-4.18(0.05)\\

\hline
\end{tabular}
\end{threeparttable}
\end{footnotesize}
\end{table*}

\begin{table}[hbt!]
\caption{Standard deviations $\sigma$ of TPA-SHF-SLy7, CPPM-Guo2013, UDLP and the new Geiger-Nuttall law.}
\label{tab3}
\renewcommand\arraystretch{1.8}
\setlength{\tabcolsep}{4.0mm}
\begin{footnotesize}
\begin{threeparttable}
\begin{tabular}{ccccc}
\hline
Model&SLy7&CPPM&UDLP&NGNL\\
\hline
$\sigma$&0.3969&0.5198&0.6063&0.4524\\
\hline
\end{tabular}
\end{threeparttable}
\end{footnotesize}
\end{table}

In the following, we use TPA-SHF-SLy7 to predict the proton radioactivity half-lives of 7 spherical proton emitters whose proton radioactivity is energetically allowed or has been observed but is not yet quantified in NUBASE2016 \cite{1674-1137-41-3-030001}. For comparison, CPPM-Guo2013, UDLP and the new Geiger-Nuttall law are also used to predict the proton radioactivity half-lives of these spherical proton emitters. The spin, parity and mass excesses used for the prediction are taken from NUBASE2016 \cite{1674-1137-41-3-030001}. Considering the overlapping effects on the half-lives of the proton emitters \cite{PhysRevC.96.034619}, we use simple power-law interpolation to obtain $Q_p$. It can be written as

\begin{equation}
\label{eq14}
Q_p=\Delta M-(\Delta M_d+\Delta M_p)+k(Z^\epsilon-Z^\epsilon_d), 
\end{equation}
where $\Delta M_p$, $\Delta M_d$ and $\Delta M$ are the mass excesses of the emitted proton, daughter nucleus and parent nucleus, respectively. $Z$ and $Z_d$ are the proton numbers of the parent nucleus and daughter nucleus, respectively. The effect of the atomic electrons $kZ^\epsilon$ represents the total binding energy \cite{Denisov2009}, and the term $k(Z^\epsilon-Z^\epsilon_d)$ represents the screening effect of the atomic electrons; for $Z \textless 60$, $k=13.6$eV, $\epsilon=2.408$ and for $Z \geq 60$, $k=8.7$eV, $\epsilon=2.517$ \cite{HUANG1976243}. 

We list the predicted results in Table \ref{tab4}. In this table, the first four columns present the parent nuclei, proton radioactivity energy $Q_p$, transferred minimum orbital angular momentum $l$, and spin and parity transformed from the parent to daughter nuclei, respectively. The next four columns present the logarithmic forms of the proton radioactivity half-lives calculated by TPA-SHF-SLy7 , CPPM-Guo2013, UDLP and the new Geiger-Nuttall law. The last column displays the lower limit of experimental data of the proton radioactivity half-lives taken from NUBASE2016 \cite{1674-1137-41-3-030001}. Based on the $\sigma$=0.3869 of SKT5 for 32 nuclei in the same region as the predicted proton emitters, the predicted proton radioactivity half-lives are within a factor of 2.50. In addition, it can be clearly seen from Table \ref{tab4} that the results predicted by UDLP and our model are consistent.

\begin{table*}[!h]
\caption{Predicted proton radioactivity half-lives of spherical proton emitters using CPPM-Guo2013, UDLP, NGNL, and TPA-SHF-SLy7. “()” represents uncertain spin and/or parity. “$\#$” indicate values estimated from trends in neighboring nuclides with the same N and Z parities. “m” and “n” mean assignments to excited isomeric states (defined as higher states with half-lives greater than 100 ns).}
\label{tab4}
\renewcommand\arraystretch{1.2}
\setlength{\tabcolsep}{1.5mm}
\begin{footnotesize}
\begin{tabular}{ccccccccc}
\hline
{Parent nuclei} & $Q_{p}$ (MeV) &${J_{p}^{\rm{\pi}}}\rightarrow$${J_{d}^{\rm{\pi}}}$& $l$ value&${\rm{lg}{T_{1/2}^{\rm{SLy7}}}}$(s)&${\rm{lg}{T_{1/2}^{\rm{CPPM}}}}$(s)&${\rm{lg}{T_{1/2}^{\rm{UDLP}}}}$(s) &${\rm{lg}{T_{1/2}^{\rm{NGNL}}}}$(s)&${\rm{lg}{T_{1/2}^{\rm{exp}}}}$(s)\\
 \hline
 $^{146}$Tm$^n$&1.144&$(10^+)\rightarrow11/2^-\#$&5&-0.05&0.145&0.069&-0.206&\\
$^{164}$Ir$^m$&1.837&$(9^+)\rightarrow7/2^-$&5&-4.368&-4.326&-3.895&-4.204&\\
$^{165}$Ir &1.557&$1/2^+\#\rightarrow0^+$&0&-6.098&-5.861&-5.236&-5.6&\\
$^{169}$Ir$^m$&0.781&$(11/2^)\#\rightarrow0^+$&5&8.687&8.868&7.713&8.065&\\
$^{169}$Au &1.947&$1/2^+\#\rightarrow0^+$&0&-8.438&-8.246&-7.378&-7.478&\\
$^{171}$Ir$^m$&0.402&$11/2^-\#\rightarrow0^+$&5&19.573&20.04&20.896&21.952&\\
$^{172}$Au &0.877&$(2^-)\rightarrow7/2^-\#$&2&4.314&4.543&3.903&3.983&$>$0.146\\
$^{172}$Au$^m$&0.627&$(9^+)\#\rightarrow13/2^+$&2&10.802&11.207&9.834&9.678&$>$-0.260\\

\hline
\end{tabular}
\end{footnotesize}
\end{table*}

\begin{figure*}[h]
\begin{minipage}{0.3\linewidth}
\centerline{\includegraphics[width=6cm]{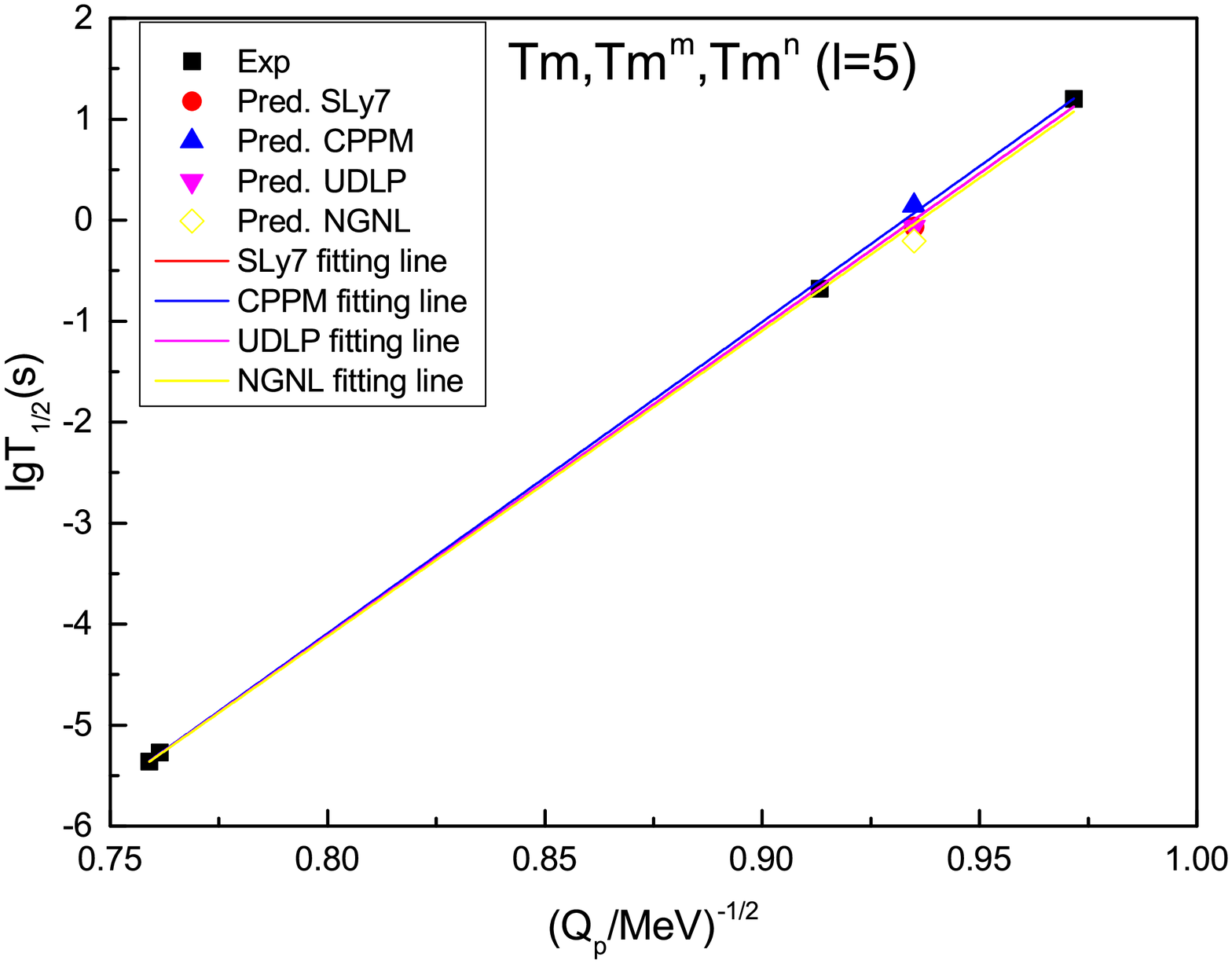}}
\centerline{The case of Tm}
\end{minipage}
\hfill
\begin{minipage}{0.3\linewidth}
\centerline{\includegraphics[width=6cm]{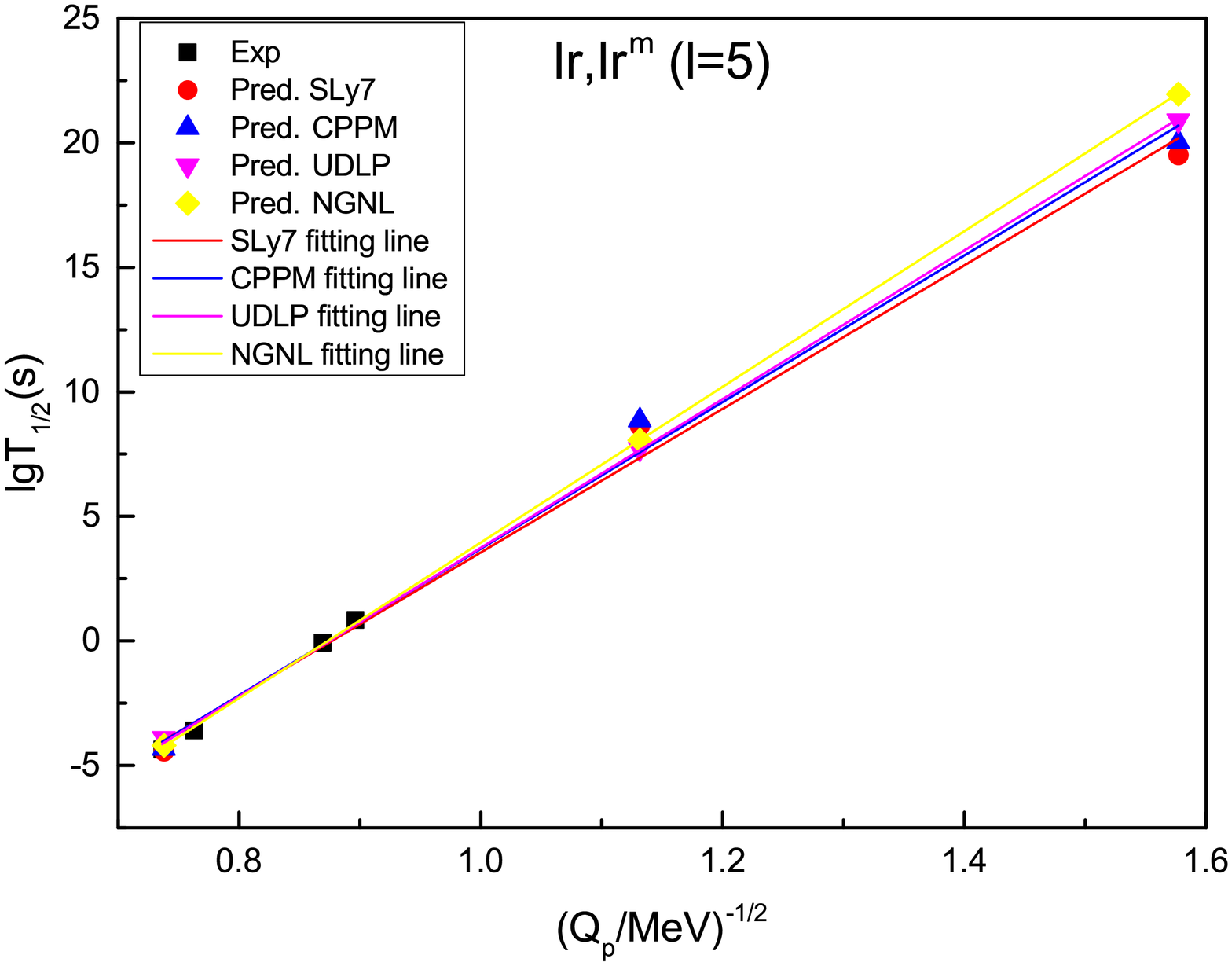}}
 \centerline{The case of Ir}
\end{minipage}
\hfill
\begin{minipage}{0.3\linewidth}
\centerline{\includegraphics[width=6cm]{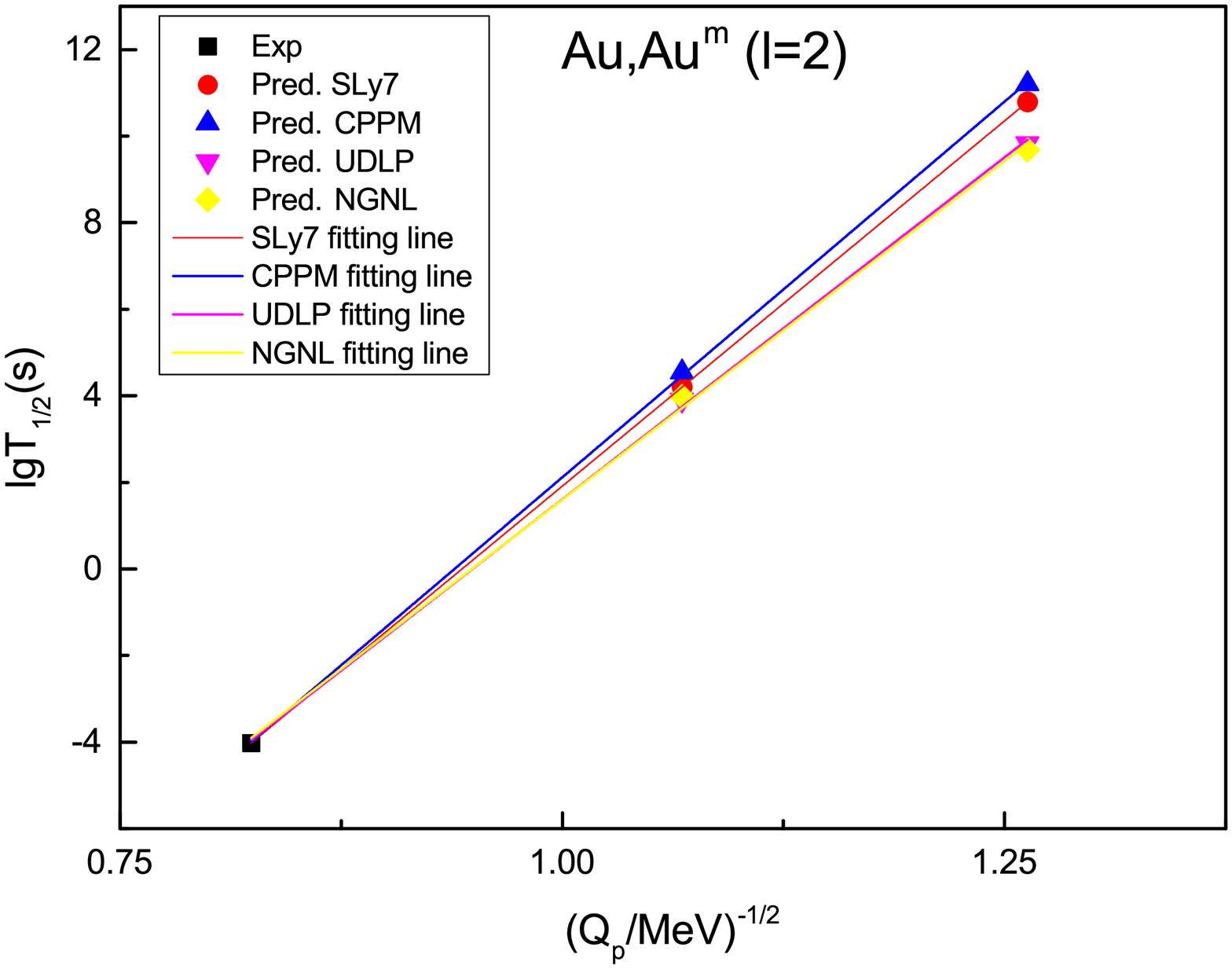}}
\centerline{The case of Au}
\end{minipage}
\caption{(Color online) Linear relationships between $Q_p^{\rm{-1/2}}$ and the theoretical spherical proton radioactivity half-life predicted by different models.}
\label{fig4}
\end{figure*}

In 1911, Geiger and Nuttall found that there is a significant correlation between the half-life of $\alpha$ decay and the $Q$ value of emitted particles \cite{Geiger}. The Geiger-Nuttall law can be extended to nonzero angular momentum \cite{QI2019214}. To test our predictions, we plot the linear relationships between ${\rm{lg}{T_{1/2}^{\rm{SLy7}}}}$, ${\rm{lg}{T_{1/2}^{\rm{CPPM}}}}$, ${\rm{lg}{T_{1/2}^{\rm{UDLP}}}}$, and ${\rm{lg}{T_{1/2}^{\rm{NGNL}}}}$ and the $Q_p^{\rm{-1/2}}$ of parent nuclei $^{170, 172}$Au, $^{172}$Au$^{m}$, $^{164}$Ir, $^{164, 165, 166, 167, 169}$Ir$^{m}$, $^{144, 145, 147, 146}$Tm$^{m}$ and $^{146}$Tm$^{n}$, in Figure \ref{fig4}. Each kind of isotope has the same angular momentum $l$. In this figure, the black square represents the experimental data of proton radioactivity, and the other different colors represent the predicted results obtained by different theoretical models. We can clearly see that ${\rm{lg}{T_{1/2}^{\rm{SLy7}}}}$ is linearly dependent on $Q_p^{\rm{-1/2}}$. This result proves that our predictions are credible.

\section{Summary}
\label{section 4}

In summary, we present a systematic study of the proton radioactivity half-lives in spherical nuclei within the framework of the two-potential approach plus Skyrme-Hartree-Fock model. 115 Skyrme interactions have been applied to calculate proton radioactivity half-lives. We found that all these Skyrme interaction can relatively well reproduce the experimental data with the TPA-SHF approach. In particular, the predictions of the SLy7 interaction gives the lowest r.m.s. deviation in the description of are in the best agreement with the experimental data. We also used this model with the Skyrme interaction of SLy7 to predict the proton radioactivity half-lives of 8 nuclei in the same region. The experimental values differ from our calculated half-lives by a factor of 2.50 on average. This work can be used as a reference for future research on proton radioactivity.\\

\noindent
This work is supported in part by the National Natural Science Foundation of China (Grant No. 11205083, No. 11505100, and No.11905302), the construct program of the key discipline in Hunan province, the Research Foundation of Education Bureau of Hunan Province, China (Grant No. 18A237 and No. 18B263), the Natural Science Foundation of Hunan Province, China (Grant No. 2015JJ3103, No. 2015JJ2123 and No.2018JJ2321), the Innovation Group of Nuclear and Particle Physics in USC, the Shandong Province Natural Science Foundation, China (Grant No. ZR2015AQ007) and the Opening Project of Cooperative Innovation Center for Nuclear Fuel Cycle Technology and Equipment, University of South China (Grant No. 2019KFZ10).

%

\begin{thebibliography}{}
%
%

\bibliographystyle{unsrt}


\bibitem{JACKSON1970281} K. P. Jackson, C. U. Cardinal, H. C. Evans \emph{et al.}, Phys. Lett. B \textbf{33}, 281(1970)

\bibitem{SONZOGNI20021}A. A. Sonzogni, Nucl. Data Sheets \textbf{95}, 1 (2002)

\bibitem{PhysRevC.85.011303}C. Qi, D. S. Delion, R. J. Liotta \emph{et al.}, Phys. Rev. C \textbf{85}, 011303 (2012)

\bibitem{Zdeb2016}A. Zdeb, M. Warda, C. M. Petrache \emph{et al.}, Eur. Phys. J. A \textbf{52}, 323 (2016)

\bibitem{Deng2019}J. G. Deng, X. H. Li, J. L. Chen \emph{et al.}, Eur. Phys. J. A \textbf{55}, 58 (2019)

\bibitem{BLANK2008403}B. Blank, M. Borge, Prog. Part. Nucl. Phys. \textbf{60}, 403 (2008)


\bibitem{Delion2006}D. S. Delion, R. J. Liotta, R. Wyss, Phys. Rep. \textbf{424}, 113 (2006)

\bibitem{Yi_Bin_2010} Y. B. Qian, Z. Z. Ren, D. D. Ni, Z. Q. Sheng, Chin. Phys. Lett. \textbf{27}, 112301 (2010)

\bibitem{PhysRevC.72.051601} D. N. Basu, P. Roy Chowdhury, C. Samanta, Phys. Rev. C \textbf{72}, 051601 (2005)

\bibitem{BAO201485} X. J. Bao, H. F. Zhang, H. F. Zhang, G. Royer, J. Q. Li, Nucl. Phys. A \textbf{921}, 85 (2014)

\bibitem{FERREIRA2011508}L. S. Ferreira, E. Maglione, P. Ring, Phys. Lett. B \textbf{701}, 508 (2011)

\bibitem{BHATTACHARYA2007263}Madhubrata Bhattacharya and G. Gangopadhyay, Phys. Lett. B \textbf{651}, 263 (2011)

\bibitem{Qian_2010}Q. Tang, X. Y. Wang, Chin. Phys. Lett. \textbf{27}, 030508 (2011)

\bibitem{PhysRevC.71.014603}M. Balasubramaniam, N. Arunachalam, Phys. Rev. C \textbf{71}, 014603 (2005)

\bibitem{DONG2010198}J. M. Dong, H. F. Zhang, Y. Z. Wang, W. Zuo, J. Q. Li, Nucl. Phys. A \textbf{832}, 198 (2010)

\bibitem{PhysRevC.84.037301}D. D. Ni, Z. Z. Ren, Phys. Rev. C \textbf{84}, 037301 (2011)

\bibitem{PhysRevC.83.067302}D. D. Ni, Z. Z. Ren, Phys. Rev. C \textbf{83}, 067302 (2011)

\bibitem{PhysRevC.56.1762}Sven \AA{}berg, B. Semmes Paul, Witold Nazarewicz, Phys. Rev. C \textbf{56}, 1762 (1997)

\bibitem{Chen_2019} J. L. Chen, X. H. Li, J. H. Cheng, J. G. Deng, X. J. Wu, J. Phys. G: Nucl. Part. Phys. \textbf{46}, 065107 (2019)




\bibitem{PhysRevLett.59.262}S. A. Gurvitz, G. Kalbermann, Phys. Rev. Lett. \textbf{59}, 262 (1987)




\bibitem{Sun_2017}X. D. Sun, X. J. Wu, B. Zheng \emph{et al.}, Chin. Phys. C \textbf{41}, 014102 (2017)

\bibitem{PhysRevC.94.024338}X. D. Sun, P. Guo, X. H. Li, Phys. Rev. C \textbf{94}, 024338 (2016)

\bibitem{PhysRevC.93.034316}X. D. Sun, P. Guo, X. H. Li, Phys. Rev. C \textbf{93}, 034316 (2016)

\bibitem{PhysRevC.95.014319}X. D. Sun, C. Duan, J. G. Deng \emph{et al.}, Phys. Rev. C \textbf{95}, 014319 (2017)

\bibitem{PhysRevC.95.044303}X. D. Sun, J. G. Deng, D. Xiang \emph{et al.}, Phys. Rev. C \textbf{95}, 044303 (2017)

\bibitem{PhysRevC.96.024318}J. G. Deng, J. C. Zhao, D. Xiang \emph{et al.}, Phys. Rev. C \textbf{96}, 024318 (2017)

\bibitem{PhysRevC.97.044322}J. G. Deng, J. C. Zhao, P. C. Chu \emph{et al.}, Phys. Rev. C \textbf{97}, 044322 (2018)

\bibitem{Buck1992}B. Buck, A. C. Merchant, S. M. Perez, Phys. Rev. C \textbf{45}, 2247-2253 (1992)

\bibitem{PhysRevC.5.626}D. Vautherin, D. M. Brink, Phys. Rev. C \textbf{5}, 626 (1972)


\bibitem{doi:10.1063/1.2827286}J. S. Al‐Khalili, A. J. Cannon, P. D. Stevenson, AIP Conf. Proc. \textbf{961}, 66 (2007)

\bibitem{Routray2012}T. R. Routray, A. Mishra, S. K. Tripathy \emph{et al.}, Eur. Phys. J. A \textbf{48}, 77 (2012)


\bibitem{PhysRevC.85.024305}R. Chen, B. J. Cai, L. W. Chen \emph{et al.}, Phys. Rev. C \textbf{85}, 024305 (2012)

\bibitem{CHABANAT1997710}E. Chabanat, P. Bonche, P. Haensel \emph{et al.}, Nucl. Phys. A \textbf{627}, 710 (1997)

\bibitem{Zhang2018}Z. Zhang, C. M. Ko, Phys. Rev. C \textbf{98}, 054614 (2018)

\bibitem{PhysRevC.82.059902}V. Y. Denisov, A. A. Khudenko, Phys. Rev. C \textbf{82}, 059902 (2010)

\bibitem{doi:10.1063/1.531270}J. Morehead James, J. Math. Phys. \textbf{36}, 5431 (1995)


\bibitem{1674-1137-41-3-030001}G. Audi, F. G. Kondev, M. Wang \emph{et al.}, Chin. Phys. C \textbf{41}, 030001 (2017)


\bibitem{1674-1137-41-3-030002}W. J. Huang, G. Audi, M. Wang \emph{et al.}, Chin. Phys. C \textbf{41}, 030002 (2017)

\bibitem{1674-1137-41-3-030003}M. Wang, G. Audi, F. G. Kondev \emph{et al.}, Chin. Phys. C \textbf{41}, 030003 (2017)





\bibitem{CHABANAT1998231}E. Chabanat, P. Bonche, P. Haensel \emph{et al.}, Nucl. Phys. A \textbf{635}, 231 (1998)
\bibitem{PhysRevC.33.335} J. Friedrich, P. G. Reinhard, Phys. Rev. C \textbf{33}, 335 (1986) 
\bibitem{TONDEUR1984297}F. Tondeur, M. Brack, M. Farine, J. M. Pearson, Nucl. Phys. A \textbf{420}, 297 (1984)
\bibitem{PhysRevC.70.044309} M. Samyn, S. Goriely, M. Bender, J. M. Pearson, Phys. Rev. C \textbf{70}, 044309 (2004)
\bibitem{m1}M. Rayet, M. Arnould, G. Paulus, F. Tondeur, Astron. Astrophys. \textbf{116}, 183 (1982)
\bibitem{VANGIAI1981379} Van Giai Nguyen, H. Sagawa, Phys. Lett. B \textbf{106}, 379 (1981) 
\bibitem{PhysRevC.66.014303}J. Margueron, J. Navarro, V. G. Nguyen, Phys. Rev. C \textbf{66}, 014303 (2002)
\bibitem{PhysRevC.72.014310} B. K. Agrawal, S. Shlomo, Au V Kim, Phys. Rev. C \textbf{72}, 014310 (2005) 
\bibitem{GORIELY2001311} S. Goriely, F. Tondeur, J. M. Pearso, Atom. Data Nucl. Data \textbf{77}, 311 (2001) 
\bibitem{PhysRevC.62.024308}F. Tondeur, S. Goriely, J. M. Pearson, M. Onsi, Phys. Rev. C \textbf{62}, 024308 (2000)
\bibitem{PhysRevC.64.027301} J. M. Pearson, S. Goriely, Phys. Rev. C \textbf{64}, 027301 (2001)
\bibitem{PhysRevC.40.2834} L. Bennour, P. H. Heenen, P. Bonche, J. Dobaczewski, H. Flocard, Phys. Rev. C \textbf{40}, 2834 (1989) 
\bibitem{PhysRevC.50.460} M. Onsi, H. Przysiezniak, J. M. Pearson, Phys. Rev. C \textbf{50}, 460 (1994) 
\bibitem{REINHARD1995467} P. G. Reinhard, H. Flocard, Nucl. Phys. A. \textbf{584}, 467 (1995)
\bibitem{PhysRevLett.102.152503} S. Goriely, N. Chamel, J. M. Pearson, Phys. Rev. Lett. \textbf{102}, 152503 (2009)
\bibitem{PhysRevC.68.054325}S. Goriely, M. Samyn, M. Bender, J. M. Pearson, Chin. Phys. C \textbf{41}, 030003 (2002)
\bibitem{PhysRevC.53.740} W. Nazarewicz, J. Dobaczewski \emph{et al}, Phys. Rev. C \textbf{53}, 740 (1996) 
\bibitem{GORIELY2006279} S. Goriely, M. Samyn, J. M. Pearson, Nucl. Phys. A \textbf{773}, 279 (2006) 
\bibitem{BARTEL198279}J. Bartel, P. Quentin, M. Brack, C. Guet, H-B Håkansson, Nucl. Phys. A \textbf{386}, 79 (1982)
\bibitem{PhysRevC.75.064312} S. Goriely, M. Samyn, J. M. Pearson, Phys. Rev. C \textbf{75}, 064312 (2007) 
\bibitem{PhysRevC.66.024326}S. Goriely, M. Samyn, P. H. Heenen, J. M. Pearson, F. Tondeur, Phys. Rev. C \textbf{66}, 024326 (2002)
\bibitem{DUTTA1986374} A. K. Dutta, J. P. Arcoragi, J. M. Pearson, R. H. Behrman, M. Farine, Nucl. Phys. A \textbf{454}, 374 (1986) 
\bibitem{PEARSON2000163} J. M. Pearson, R. C. Nayak, Nucl. Phys. A \textbf{668}, 163 (2000) 
\bibitem{SAMYN2002142} M. Samyn, S. Goriely, P. H. Heenen, J. M. Pearson, F. Tondeur, Nucl. Phys. A \textbf{700}, 142 (2002) 
\bibitem{PEARSON19911} J. M. Pearson, Y. Aboussir, A. K. Dutta, R. C. Nayak, M. Farine, F. Tondeur, Nucl. Phys. A \textbf{528}, 1 (1991) 
\bibitem{KOHLER1976301} H. S. Köhler, Nucl. Phys. A \textbf{258}, 301 (1976) 
\bibitem{GORIELY2001349} S. Goriely, M. Pearson, F. Tondeur, Nucl. Phys. A \textbf{688}, 349 (2001) 
\bibitem{CHAMEL200872}N. Chamel, S. Goriely, J. M. Pearson, Nucl. Phys. A \textbf{812}, 72 (2008)
\bibitem{PhysRevC.68.031304}B. K. Agrawal, S. Shlomo, V. Kim Au, Astron. Astrophys. \textbf{68}, 031304 (2003)
\bibitem{KRIVINE1980155}H. Krivine, J. Treiner, O. Bohigas, Nucl. Phys. A \textbf{336}, 155 (1980)
\bibitem{BEINER197529} M. Beiner, H. Flocard, Van Giai Nguyen, P. Quentin, Nucl. Phys. A \textbf{238}, 29 (1975)
\bibitem{PhysRevC.58.2099} B. A. Brown, W. A. Richter, Phys. Rev. C \textbf{58}, 2099 (1998) 
\bibitem{ABOUSSIR1992155} Y. Aboussir, J. M. Pearson, A. K. Dutta, F. Tondeur, Nucl. Phys. A \textbf{549}, 155 (1992) 
\bibitem{PhysRevC.52.2254} R. C. Nayak, J. M. Pearson, Phys. Rev. C \textbf{52}, 2254 (1995) 
\bibitem{PhysRevC.21.2076} M. J. Giannoni, P. Quentin, Phys. Rev. C \textbf{21}, 2076 (1980) 
\bibitem{PhysRevC.60.014316} P. G. Reinhard, D. J. Dean \emph{et al.}, Phys. Rev. C \textbf{60}, 014316 (1999)
\bibitem{PhysRevC.77.031301} S. Goriely, J. M. Pearson, Phys. Rev. C \textbf{77}, 031301 (2008) 
\bibitem{KO1974269} C. M. Ko, H. C. Pauli, M. Brack, G. E. Brown, Nucl. Phys. A \textbf{236}, 269 (1974) 
\bibitem{PhysRevC.73.014313} L. G. Cao, U. Lombardo, C. W. Shen, Giai Nguyen Van, Phys. Rev. C \textbf{73}, 014313 (2006)
\bibitem{Treiner_1976} J. Treiner, H. Krivine, J. Phys. G: Nucl. Part. Phys. \textbf{2}, 285 (1976) 
\bibitem{PETHICK1995675} C. J. Pethick, D. G. Ravenhall, C. P. Lorenz, Phys. Rev. C \textbf{584}, 675 (1995) 
\bibitem{PhysRevLett.74.3744} M. M. Sharma, G. Lalazissis, J. K\"onig, P. Ring, Phys. Rev. Lett. \textbf{74}, 3744 (1995) 












\bibitem{PRL2012}C. Qi, F. R. Xu, R. J. Liotta \emph{et al.}, Phys. Rev. Lett.\textbf{103}, 072501 (2009)

\bibitem{Chen20192}J. L. Chen, J. Y. Xu, J. G. Deng \emph{et al.}, Eur. Phys. J. A \textbf{55}, 214 (2019)

\bibitem{PhysRevC.96.034619}K. P. Santhosh, I. Sukumaran, Phys. Rev. C \textbf{96}, 034619 (2017)

\bibitem{Denisov2009}V. Y. Denisov, A. A. Khudenko, Atom. Data Nucl. Data \textbf{95}, 815–835 (2009) 

\bibitem{HUANG1976243}K. N. Huang, M. Aoyagi, M. H. Chen \emph{et al.}, Atom. Data Nucl. Data \textbf{18}, 243 (1976)

\bibitem{Geiger}H. Geiger, J. M. Nuttall, Philos. Mag. \textbf{22}, 613(1911) 

\bibitem{QI2019214}C. Qi, R. Liotta, R. Wyss, Prog. Part. Nucl. Phys. \textbf{105}, 214(2019) 

\end{thebibliography}
%

\end{document}